\documentclass[referee]{aa} 
\usepackage[latin1]{inputenc}
\usepackage{natbib}
\usepackage{theorem,algorithm,algorithmic}
\usepackage{amssymb,amsmath}
\usepackage{psfig,epsfig,amssymb,alltt}
\usepackage{latexsym}
\usepackage{graphicx}
\graphicspath{{PS/}}
\usepackage{psfig}
\psfigurepath{./PS:.}
\usepackage[english]{babel}
\usepackage{indentfirst}
\psfigurepath{./PS}
\vspace{1cm}

\usepackage{epsfig}

\psfigurepath{./PS}

\newcommand{\be}{\begin{eqnarray}}
\newcommand{\ee}{\end{eqnarray}}

\newcommand{\bitem}{\begin{itemize}}
\newcommand{\eitem}{\end{itemize}}

\authorrunning{Starck \textit{et al.}}
\titlerunning{Polarized wavelets and curvelets}

\title{Polarized wavelets and curvelets on the sphere} 

\author{J.-L.~Starck, Y.~Moudden and J.~Bobin}
\institute{
Laboratoire AIM (UMR 7158), CEA/DSM-CNRS-Universit\'e Paris Diderot,   
IRFU, SEDI-SAP, Service d'Astrophysique,  Centre de Saclay, 
F-91191 Gif-Sur-Yvette cedex, France 
}

\offprints{jstarck@cea.fr}

\date{\today}

\begin{document}

\abstract{
The statistics of the temperature anisotropies in the primordial cosmic microwave background radiation field provide a wealth of information for cosmology and for estimating cosmological parameters. 
An even more acute inference should stem from the study of maps of the polarization state of the CMB radiation. 
Measuring the extremely weak CMB polarization signal requires very sensitive instruments. The full-sky maps of both temperature and polarization anisotropies of the CMB to be delivered by the upcoming Planck Surveyor satellite experiment are hence being awaited with excitement. 
Multiscale methods, such as isotropic wavelets, steerable  wavelets, or  curvelets, have been proposed in the past to analyze the CMB temperature map. In this paper, 
we  contribute to enlarging the set of available transforms for polarized data on the sphere.  We describe a set of new  multiscale decompositions for polarized data
on the sphere, including decimated and undecimated  Q-U or E-B wavelet transforms  and Q-U or E-B curvelets.
The proposed transforms are invertible and so allow for applications in data restoration and denoising.
}

\maketitle


\section{Introduction}
 \label{sec:intro}
The statistical analysis of the slight intensity fluctuations in the primordial cosmic microwave background radiation field, for which evidence was found for the first time in the early 1990's in the observations made by COBE~\citep{gauss:smoot92},  is a major issue in modern cosmology  as these are strongly related to  the cosmological scenarios describing the properties and evolution of our Universe. In the Big Bang model, the observed CMB anisotropies are an imprint of primordial fluctuations in baryon-photon density from a time when the temperature of the Universe was high enough above  3000~K for matter and radiation to be tightly coupled. At that time, the attraction of  gravity and the repulsive radiation pressure were opposed,  thus generating so-called acoustic oscillations in the baryon-photon fluid, causing peaks and troughs to appear in the power spectrum of the spatial anisotropies of the CMB. With the Universe cooling down as it expanded, matter and radiation finally decoupled. Photons were set free in a nearly transparent Universe, while the density fluctuations collapsed under the effect of gravity into large scale structures such as galaxies or clusters of galaxies. Due to the expansion of the Universe, the CMB photons are now observed in the microwave range but are still distributed according to an almost perfect black body emission law. Another major result was the measurement of the polarization state and anisotropies of the CMB radiation field by DASI~\citep{dasi}. Only a fraction of the total CMB radiation is polarized so that extremely sensitive instruments are needed. Polarization of the CMB radiation is a consequence of the Thomson scattering of photons on electrons. But for the outgoing population of photons to be polarized, the radiation incident on the scatterer needs to be anisotropic and have a quadrupole moment. The statistics of the CMB polarization anisotropies are also a source of information for cosmology. Inference of  cosmological parameters from the joint statistics of the  CMB  anisotropies should benefit from both the complementarity  and the redundancy of the information carried by the additional measurement of CMB polarization. Hence the full-sky maps with unprecedented sensitivity and angular resolution of both temperature and polarization anisotropies of the CMB to be delivered by the upcoming Planck Surveyor satellite experiment are awaited with excitement.

\subsection*{Multiscale transforms on the sphere}

Wavelets have been used  in many studies of CMB data analysis, especially for non-Gaussianity detection  \citep{gauss:aghanim99,hobson99,gauss:vielva04,starck:jin05,starck:sta03_1,vielva06, McEwen08}. Continuous  wavelet transforms on the sphere~\citep{wave:antoine99,wave:tenerio99,wave:cayon01,wave:holschneider96} are the most used transforms.
Recently,  \citet{starck:sta05_2}  proposed an invertible isotropic undecimated wavelet transform (UWT) on the sphere based on 
spherical harmonics. A similar wavelet construction has been published  in \citep{marinucci08,fay08a,fay08} using so-called "needlet filters".
 \citet{wiaux08} have  also proposed an algorithm which allows us to reconstruct an image from its steerable wavelet transform.
Since reconstruction algorithms are available,  these new tools can be used for other applications than non-Gaussianity detection, 
such as denoising, deconvolution, component separation \citep{starck:yassir05,bobin-gmca-cmb,delabrouille08} or inpainting \citep{inpainting:abrial06,starck:abrial08}.
Other  multiscale  transforms on the sphere such as ridgelets and curvelets have been developed  \citep{starck:sta05_2}, and are well adapted to detect anisotropic features.
It has been shown that such constructions are very useful for the detection of cosmic strings \citep{starck:sta03_1,starck:jin05,hammond08}.
More generally, each transform can be characterized by a given dictionary $\Phi = \{ \phi_1, ..., \phi_K \} $, and the transformation consists in representing the input data $y$ as a linear combination  of the dictionary elements: $y =  \Phi \alpha = \sum_k \phi_k \alpha_k$, where $\alpha$ are the coefficients. In the case of a wavelet transform, the dictionary is composed of wavelet functions, and $\alpha$ are the wavelet coefficients. Depending on the content of the data, a given dictionary may be more adapted to detect the signal of interest.
A dictionary is considered as well designed  for a class of signals if the transformation of any of these signals leads to a sparse representation, i.e. if most of the coefficients 
are equal or close to zero, while only few coefficients have a high amplitude. 
If the morphological signature of the features to be detected  are not known, we should not restrict ourselves to analyze the data with wavelets or curvelets only, but
rather use all available tools \citep{starck:sta03_1}.

\bigskip

In this paper, we  describe a set of new multiscale decompositions for polarized data on the sphere, including decimated and undecimated  Q-U or E-B wavelet transforms, and Q-U or E-B curvelets. Choosing the right transform will depend on the application and on the typical structures occuring in the data to be analyzed. Indeed, as illustrated next, the different transforms are associated with different waveform dictionaries.     
Section~\ref{sec:polar} presents an orthogonal decomposition for polarized data on the sphere.
Section~\ref{sec:modphase} introduces a new decomposition where the modulus and the phase are processed independently. 
Sections \ref{sec:pol_iwt} and \ref{sec:pol_cur} extend the isotropic wavelet transform and the curvelet transform on the sphere to the case of polarized data, and
section~\ref{sec:pol_eb} introduces two other wavelet and curvelet decompositions which are based on the E/B mode separation.
The proposed transforms are invertible and so allow for applications in data restoration and denoising. 
Section~\ref{sec:pol_denoising} reports on denoising experiments using these new polarized transforms, thus demonstrating their usefulness in practice. 


\section{Orthogonal representation of polarized data}
\label{sec:polar}
\subsection{Introduction}
\begin{figure*}[htb]
\caption{Q-U orthogonal Wavelet Transform.}
\label{fig_qu_owt_trans}
\end{figure*}
Full-sky CMB polarization data, as expected from the upcoming Planck experiment, consists of measurements of the Stokes parameters so that in addition to the temperature $T$ map, $Q$ and $U$ maps are given as well. The fourth Stokes parameter commonly denoted $V$ is a measure of circular polarization. In the case of CMB which is not expected to have circularly polarized anisotropies, $V$ vanishes. The former three quantities, $T$, $Q$ and $U$ then fully describe the linear polarization state of the CMB radiation incident along some radial line of sight : $T$ is the total incoming intensity, $Q$ is the difference between the intensities transmitted by two perfect orthogonal polarizers the directions of which define a reference frame in the tangent plane, and $U$ is the same as $Q$ but with polarizers rotated 45 degrees in that tangent plane. Clearly, $Q$ and $U$ are not invariant through a rotation of angle $\phi$ of the local reference frame around the line of sight. In fact, it is easily shown that~:
\begin{eqnarray}
Q ' = & \cos (2 \phi) Q + \sin(2 \phi) U \\ \nonumber
U ' = & \cos (2 \phi) U - \sin(2 \phi) Q 
\end{eqnarray}
which can also be written $Q' \pm i U' = e^{\mp i\phi}  ( Q \pm i U )$ which by definition expresses the fact that the quantities $Q \pm i U$ are spin-2 fields on the sphere. 
The suitable generalization of the Fourier representation for such fields is the spin-2 spherical harmonics basis denoted $_{\pm 2}Y_{\ell m}$, in which we can expand~: 
\begin{eqnarray}\label{QU}
Q \pm i U  = \sum_{\ell, m}  { _{\pm 2}a_{\ell m}}   {_{\pm 2}Y_{\ell m} }
\end{eqnarray} 
 
 \begin{figure*}[htb]
\centerline{
 \vbox{
 }
 }
\caption{ \textit{Top~:} examples of backprojections of Q-wavelet coefficients. \textit{Bottom~:}  examples of backprojections of U-wavelet coefficients.}
\label{fig_qu_owt_back}
\end{figure*}

 \subsection{Multiscale Representation}
The easiest way to build a multiscale transform for polarized data is to use the Healpix\footnote{http://healpix.jpl.nasa.gov} representation  \citep{pixel:healpix}, and to apply a bi-orthogonal wavelet transform on each face of the Healpix map, separately for $Q$ and $U$.
Fig.~\ref{fig_qu_owt_trans} shows the flow-graph of this Q-U orthogonal wavelet transform (QU-OWT). Recall that the base resolution of the Healpix representation divides the sphere into twelve curvilinear quadrilateral faces of equal area placed on three rings around the poles and equator. Each face is subsequently divided into $nside^{2}$ pixels of exactly equal surface but with varying shape. It follows that $Q$ and $U$  are 
reconstructed at position $k$ from their wavelet coefficients $w_{j,p}^Q$, $w_{j,p}^U$, $c^Q_{J,p}$ and $c^U_{J,p}$ according to~: 
  \begin{eqnarray}
 Q_k &  =  &  \sum_p c^Q_{J,p} \phi_{j,k}(p) + \sum_p \sum_{j=1}^J \psi_{j,k}(p)w_{j,p}^Q \\ \nonumber
 U_k  & =  &  \sum_p c^U_{J,p} \phi_{j,k}(p) + \sum_p \sum_{j=1}^J \psi_{j,k}(p)w_{j,p}^U
 \end{eqnarray}
which can also be written as:
 \begin{eqnarray}
(Q \pm iU)_k & =  & \sum_p (c^Q_{J,p} \pm i c^U_{J,p}) \phi_{j,k}(p) + \nonumber \\
                      &  &  \sum_p \sum_{j=1}^J \psi_{j,k}(p) ( w_{j,p}^Q \pm i w_{j,p}^U )
\label{eq_qu_owt}
\end{eqnarray}
This wavelet transform is not redundant \textit{i.e.} the decomposition has the same number of coefficients as the input data, and it is invertible so that the $Q$ and $U$ maps can be reconstructed exactly.\\

When we apply such a decomposition, we implicitly use a  dictionary $\Phi$ on which we project the data.
As discussed previously, the shape of the dictionary elements, also called atoms, is very important to have  an efficient analysis of the data. In the case of polarized data, it is not straightforward to imagine these shapes from Eq.~\ref{eq_qu_owt}.  In order to visualize them, we can perform a backprojection \textit{i.e.}
we apply the inverse wavelet transform to sets of wavelet coefficients where only one coefficient is different from zero. Repeating the same experiment, changing only the 
scale and position of the non-zero coefficient allows us to view the different atoms in the dictionary related to the  QU-OWT transform that we use. 
Fig.~\ref{fig_qu_owt_back} shows examples of backprojections of $Q$ wavelet coefficients (top) and backprojections of $U$ wavelet coefficients (bottom).
The shapes of the individual atoms do not look close to the astronomical patterns we would expect in our data. Therefore, this decomposition may not be optimal to analyze polarized astronomical data, although this would need to be confirmed in practice. The following sections describe other polarized wavelet transforms with different morphological properties.

\section{Module-phase non linear multiscale transform}
\label{sec:modphase}

\subsection{Introduction}
Given a polarized map in the standard Q-U representation, consider a different point of view and define the modulus $M$ and phase $P$ maps as follows~:
%
\begin{eqnarray}
\forall k,\,\,\, M_k & = & \sqrt{Q_k^2 +  U_k^2} \\
\forall k,\,\,\, P_k & = & \exp(i \theta_k) \mbox{ where } tan(\theta_k) = U_k/Q_k 
\end{eqnarray}
Because the smoothness of the $Q$ and $U$ maps should result in some smoothness of the modulus map $M$ and the phase map $P$, one may consider devising a multiscale modulus/phase decomposition of the 
spin 2 field ${\mathcal{V}} = \left[ Q \, U\right]$.\\
The specificity of the modulus/phase decomposition of 
$\mathcal{V}$ is twofold~: i) the modulus field is 
non-negative and ii) the phase field takes its values on the unit circle  $S^1$. 
Recently,  \citep{rahman05} introduced a multiscale analysis technique for manifold valued data that will be described in the following paragraph. We then define the modulus/phase (MP) multiscale transform as follows~:
\vspace{.1cm}
\begin{center}
\begin{minipage}[b]{0.85\linewidth}
\footnotesize{
\begin{enumerate}
\item Apply a classical multiscale transform (\textit{i.e.} wavelets) to the modulus map $M$.
\item Apply the multiscale analysis technique for manifold valued data described in \citep{rahman05} to the phase map $P$. 
\end{enumerate}}
\end{minipage}
\end{center}
\vspace{.1cm}

\subsection{Decimated MP-multiscale transform}

\begin{figure*}[htb]
\centerline{
 \vbox{
 }
 }
\caption{Examples of MP-multiscale coefficients backprojection.}
\label{fig_modphase_back}
\end{figure*}
Let us provide some essential notation~: we assume that the entries of the phase map $P$ lie in a manifold $\mathcal{M}$ (\textit{e.g.} $\mathcal{M}\equiv S^1$). According to \citep{rahman05}, 
take $p_0,p_1 \in \mathcal{M}$
and define $Log_{p_0}(p_1)$ as the log-map
of $p_1$ onto the tangent space $\mathcal{T}_{p_0}$ of $\mathcal{M}$ at $p_0$. The back-projection is obtained using the inverse of the log-map 
$Exp_{p_0}$. \footnote{In differential geometry, the Exp map and Log map are generalizations of the usual exponential and logarithm function. In this paper, the manifold $\mathcal{M}$ is a Riemannian manifold. In that case, the Exp map at point $p_0$, $Exp_{p_0}(s)$ is the map which takes a vector $s$ of the tangent space of $\mathcal{M}$ at $p_0$ and provides the point $p_1$ by travelling along the geodesic starting at $p_0$ in the direction s.}\\
For instance, if we choose $\mathcal{M} \equiv S^1$ then $p_0 = \exp(i \theta_0)$ and $p_1 = \exp(i \theta_1)$. The $Exp_{p_0}$ and $Log_{p_0}$ maps are then defined as follows~:
\begin{eqnarray}
\forall p_1 \in S^, \,\,\,  Log_{p_0} (p_1) & = & \theta_1 - \theta_0 \\
\forall s \in \mathbb{R} \,\,\, Exp_{p_0} (s) & = & exp(i(\theta_0 + s))
\end{eqnarray}
The multiscale transform for manifold valued data introduced in \citep{rahman05} is equivalent to a two-step interpolation-refinement scheme similar to the lifting scheme described in~\citep{wave:sweldens98}. The wavelet coefficients and low pass approximation pixels are then computed as follows at each scale $j$ and pixel $k$~:
\begin{eqnarray}\label{eq:mani}
w_{j+1,k}^P & = & Log_{c_{j,2k+1}^P}\left(\mathcal{P}(c_{j,2k}^P)\right)   \\
c_{j+1,k}^P & = & Exp_{c_{j,2k}^P} ( -\mathcal{U}(w_{j+1,k}^P))
\label{eq:mani2}
\end{eqnarray}
The wavelet coefficient $w_{j+1,k}^P$ at pixel $k$ and scale $j$ is the projection of its prediction/interpolation $\mathcal{P}(c_{j,2k}^P)$ onto the tangent space $\mathcal{T}_{c_{j,2k+1}^P}$ of $\mathcal{M}$ at $c_{j,2k+1}^P$. The low pass approximation $c_{j+1,k}^P$ at scale $j+1$ is computed by updating $c_{j,2k}^P$ from the wavelet coefficient $w_{j+1,k}^P$.\\
The main advantage of this scheme is its ability to capture local regularities while guaranteeing the low pass approximation to belong to the manifold $\mathcal{M}$. Indeed, the wavelet coefficient $w_{j+1,k}^P$ at pixel $k$ and scale $j+1$ is computed as the $Exp$ map at $c_{j,2k+1}^P$ of an approximation $\mathcal{P}(c_{j,2k}^P)$ of $c_{j,2k}^P$.\\
Note also that even if the definitions of the $Exp_{p_0}$ and $Log_{p_0}$ maps involve the absolute phase $\theta(k)$ (\textit{i.e.} $tan(\theta(k)) = U_k/Q_k$), at least they only require the computation of differences of phases values thus avoiding the explicit manipulation of an absolute phase.\\
However the non-linearity of the proposed transform is a major drawback when considering denoising and restoration applications.\\ 

\paragraph{Illustration~:\\}
In the case of polarized data, the entries of the phase map $P$ lie in $\mathcal{M} \equiv S^1$. In the following experiments, $\mathcal{P}$ and $\mathcal{U}$ are chosen such that~:
\begin{eqnarray}
w_{j+1}^P & = & Log_{c_{j,2k+1}^P}(c_{j,2k}^P)  \\
c_{j+1,k}^P & = & Exp_{c_{j,2k}^P} \left( - \frac{w_{j+1}^P}{2}\right)
\end{eqnarray}
This multiscale transform is invertible and its inverse is computed as follows~:
\begin{eqnarray}
c_{j,2k}^P & = & Exp_{c_{j+1,k}^P} \left(\frac{w_{j+1}^P}{2}\right)\\
c_{j,2k+1}^P & = & Exp_{c_{j,2k}^P}\left( w_{j+1}^P \right)   
\end{eqnarray}
The picture in Figure~\ref{fig_modphase_back} features some examples of backprojections of MP-multiscale coefficients.

\subsection{Undecimated MP-multiscale transform}
For image restoration purposes, the use of undecimated multiscale transforms has been shown to provide better results than decimated transforms~\citep{starck:book98,starck:book06}. The aforementioned modulus/phase multiscale analysis can be extended to an undecimated scheme
consisting in~: i) applying an undecimated wavelet transform to the modulus map, ii) analyzing the phase map P using an extension to the undecimated case of the multiscale transform described in \citep{rahman05}.
In that case, Equations~\eqref{eq:mani} and \eqref{eq:mani2} are replaced with the following equations~:
\begin{eqnarray}\label{eq:maniu}
c_{j+1,k}^P & = & Exp_{c_{j,k}^P} ( \mathcal{F}(c_{j,.}^P))\\
w_{j+1}^P & = & Log_{c_{j+1,k}^P}\left(c_{j,k}^P\right)  
\end{eqnarray}
where $\mathcal{F}(c_{j,.}^P) = \sum_l h_{l} Log_{c_{j,k}}\left(c_{j,k-2^jl}\right)$ with $\sum_l h_l = 1$ and $h_l > 0$. The low pass approximation $c_{j+1,k}^P$ is then computed from a linear combination (linear filter) of a neighborhood $\{c_{j,k-2^jl}\}_l$ of $c_{j,k}$ weighted by the positive scalars $\{h_l\}_l$. Note that from scale $j$ to scale $j+1$, the spatial size of the neighborhood increases by a factor $2$ which would be equivalent to downsize by a factor $2$ the band pass filter of the classical wavelet decomposition scheme.\\

\subsection{Example}
In the case of polarized data, the entries of the phase map $P$ lie in $\mathcal{M} \equiv S^1$. In the following experiments, $\mathcal{F}$ is chosen such that~:
\begin{eqnarray}
c_{j+1,k}^P & = & Exp_{c_{j,k}^P}\left(\sum_l h_{l}Log_{c_{j,k}}\left(c_{j,k-2^jl}^P\right)\right)  \\
w_{j+1,k}^P & = & Exp_{c_{j,k}^P} \left(c_{j+1,k}^P\right)
\end{eqnarray}
where~:
\begin{equation}
h_l = \left\{
\begin{array}{ccc}
0 & \mbox{ if } & l < -2 \mbox{ or } l > 2 \\
1/16 &\mbox{ if }& l=-2 \mbox{ or } l=2 \\
1/4 &\mbox{ if }& l=-1 \mbox{ or } l=1\\
3/8 &\mbox{ if }& l= 0
\end{array}
\right.
\end{equation}
This multiscale transform is invertible and its inverse is computed as follows~:
\begin{equation}
 c_{j,k}^P  =  Exp_{c_{j+1,k}^P } \left(- w_{j+1,k}^P\right)  
\end{equation}

\begin{figure*}[htb]
 \vbox{
 \centerline{
 \hbox{
 }
 }
 \centerline{
 \hbox{
 }
 }
 }
\caption{ Polarized field smoothing - \textit{top left~:} simulated synchroton emission.  \textit{top right~:} same field corrupted by additive noise.
 \textit{bottom left~:} MP-multiscale reconstruction after setting to zero all coefficients from the three first scales.
\textit{bottom right~:} MP-multiscale reconstruction after setting to zero all coefficients from  the five first scales.}
\label{fig_modphase_smoothfield}
\end{figure*}
Fig.~\ref{fig_modphase_smoothfield} top  shows a simulated polarized field of the synchrotron emission and its noisy version.
We have applied the MP-multiscale transform and we remove the first three scales
(i.e. we put all coefficients to zero) before reconstructing. The resulting image is shown on the bottom left of Fig.~\ref{fig_modphase_smoothfield}.  The bottom right of Fig.~\ref{fig_modphase_smoothfield} corresponds to the same experiment, but by removing the five first scales. We can see that the field is smoother and smoother, but respecting the large scale structure of the field. This transform will be very well suited to CMB studies where the phase is analyzed independently of the modulus, such as in~\cite{coles05,naselsky05}.


\section{Polarized Wavelet Transform using Spherical Harmonics}
\label{sec:pol_iwt}

\subsection{Isotropic Undecimated Wavelet Transform on the Sphere (UWTS) }
\label{sec:UWTS}

\begin{figure*}[htb]
\centerline{
 \hbox{
 }
 }
\caption{Q-isotropic wavelet transform backprojection (left) and U-isotropic wavelet backprojection (right).}
\label{fig_qu_iwt_back}
\end{figure*}

The undecimated isotropic transform on the sphere described in~\citep{starck:sta05_2} is similar 
in many respects to the 
usual \emph{ \`a trous} isotropic wavelet transform. It is obtained using a zonal scaling function $\phi_{l_c}(\vartheta, \varphi)$ which depends only on colatitude $\vartheta$ and is invariant with respect to a change in longitude $\varphi$. It follows that the spherical harmonic coefficients $\hat \phi_{l_c} (l,m)$ of $\phi_{l_c}$ vanish  when $m \ne 0$ which makes it simple to compute spherical harmonic coefficients $\hat c_{0}(l,m)$ of $c_0 = \phi_{l_c} * f$ where $*$ stands for convolution :
\begin{eqnarray}
 \hat c_{0}(l,m) = \widehat{\phi_{l_c} * f} (l,m) =  \sqrt{\frac{4\pi}{2l+1} }  \hat \phi_{l_c} (l,0) \hat f(l,m) 
\end{eqnarray}
A possible scaling function~\citep{starck:book98}, defined in the spherical harmonics representation, is  $\phi_{l_c}(l, m)  =  {2 \over 3} B_{3} ( {  {2 l} \over {l_{c} } } )$ where $B_{3}$ is the  cubic B-spline compactly supported over $[-2, 2]$. Denoting $\phi_{2^{-j} l_{c} }$ a rescaled version of $\phi_{l_{c}}$ with cut-off frequency $2^{-j} l_{c}$,  a multi-resolution decomposition of $f$ on a dyadic scale is obtained recursively : 
\begin{eqnarray}
c_0   & = &  \phi_{ l_{c} }  * f    \nonumber    \\
c_j    &=&   \phi_{2^{-j}  l_{c}  }  * f  =   c_{j-1} * h_{j-1} \nonumber    \\
\end{eqnarray}
where the zonal low pass filters $h_{j}$ are defined by 
\begin{eqnarray}
 \hat{H}_{j}(l,m)  & =  &  \sqrt{\frac{4\pi}{2l+1} }  \hat h_{j}(l,m)  \nonumber \\
 &  =  & \left\{
  \begin{array}{ll}
  \frac {   \hat \phi_{\frac{l_{c}}{2^{j+1}} }(l,m)   }   {  \hat  \phi_{  \frac{l_{c}}{2^{j}} }(l,m)   } & \mbox{if }  l  < \frac{ l_{c}} {2^{j+1}} \quad \textrm{and}\quad m = 0\\
0 & \mbox{otherwise } \ 
  \end{array}
  \right.
\end{eqnarray}
The cut-off frequency is reduced by a factor of $2$ at each step so that in applications where 
this is useful such as compression, the number of samples could be reduced adequately.
Using a pixelization scheme such as Healpix \citep{pixel:healpix}, this can easily be done  by dividing 
by 2 the Healpix {\it nside} parameter when computing the inverse spherical harmonics transform. 
As in the  \emph{\`a trous} algorithm, the wavelet coefficients can be defined as the difference between two consecutive resolutions, $w_{j+1}(\vartheta, \varphi) = c_{j}(\vartheta, \varphi) - c_{j+1}(\vartheta, \varphi)$. This defines a  zonal wavelet function $\psi_{l_c}$ as 
\begin{eqnarray}\label{wavelet}
\hat \psi_{\frac{l_c}{2^{j}}}(l,m) = \hat \phi_{\frac{l_c}{2^{j-1}}} (l,m)  - \hat \phi_{\frac{l_c}{2^{j}}}(l,m)
\end{eqnarray}

\begin{figure*}[htb]
\centerline{
\hbox{
}}
\caption{Simulated observations on the sphere of the polarized galactic dust emission.}
\label{fig_simu_pol_dust}
\end{figure*}

With this particular choice of wavelet function, the decomposition is readily inverted by summing the coefficient maps on all wavelet scales
 \begin{eqnarray}\label{IWT}
   f(\vartheta, \varphi) = c_{J}(\vartheta, \varphi) + \sum_{j=1}^{J} w_j(\vartheta, \varphi)
\end{eqnarray}
where we have made the simplifying assumption that $f$ is equal to $c_0$. Obviously, other wavelet functions $\psi$ could be used just as well, such as the needlet function~\citep{marinucci08}.


\begin{figure*}[htb]
\centerline{
\vbox{
 \hbox{
 }
 \hbox{
 }
  \hbox{
 }
  }
 }
\caption{QU-Undecimated Wavelet Transform of the simulated polarized map of galactic dust emission shown in figure~(\ref{fig_simu_pol_dust}).}
\label{fig_quwt_trans_dust}
\end{figure*}

\subsection{Extension to Polarized Data}
By applying the above scalar isotropic wavelet transform to each component $T$, $Q$, $U$ of a polarized map on the sphere, we have~:
\begin{eqnarray}
\label{tqu_iwt}
T(\vartheta, \varphi) & = c_{J}^T (\vartheta, \varphi)+ \sum_{j=1}^{J} w_j^T  (\vartheta, \varphi)\\ \nonumber
Q(\vartheta, \varphi) & = c_{J}^Q (\vartheta, \varphi)+ \sum_{j=1}^{J} w_j^Q (\vartheta, \varphi)\\ \nonumber
U(\vartheta, \varphi) & = c_{J}^U (\vartheta, \varphi)+ \sum_{j=1}^{J} w_j^U (\vartheta, \varphi)
\end{eqnarray}
where $c_{J}^X$ stands for the low resolution approximation to component $X$ and $w_j^X$ is the map of wavelet coefficients of that component on scale $j$. This leads to the following decomposition~:
\begin{eqnarray}
(Q \pm iU)[k] =    (c^Q_{J} \pm c^U_{J,p})[k]   +   \sum_{j=1}^J   ( w_{j}^Q \pm w_{j}^U )[k]
\label{eq_qu_rec_uwt}
\end{eqnarray}
Fig.\ref{fig_qu_iwt_back} shows the backprojection of a Q-wavelet coefficient (left) and a $U$-wavelet coefficient (right).
Fig.~\ref{fig_quwt_trans_dust} shows the undecimated isotropic polarized  wavelet transform of the dust image shown on Fig.~\ref{fig_simu_pol_dust} using six scales, \textit{i.e.} five wavelet scales and the coarse approximation.
\section{Polarized Curvelet Transform}
\label{sec:pol_cur}
 \begin{figure*}[htb]
\centerline{
\vbox{
 \hbox{
 }
 \hbox{
 }
  }
 }
\caption{Top, Q-curvelet backprojection (left)  and zoom (right). Bottom, U-curvelet backprojection (left)  and zoom. }
\label{fig_qucur_back}
\end{figure*}
The 2D ridgelet transform \citep{cur:candes99_1} was developed in an attempt to overcome some limitations inherent in former multiscale methods \emph{e.g.} the 2D wavelet, when handling smooth images with edges \textit{i.e.} singularities along smooth curves. Ridgelets are translation invariant \emph{ridge} functions with a wavelet profile in the normal direction. Although ridgelets provide sparse representations of smooth images with straight edges, they fail to efficiently handle edges along curved lines. This is the framework for curvelets which were given a first mathematical description in \citep{Curvelets-StMalo}. Basically, the curvelet dictionary is a multiscale pyramid of localized directional functions with anisotropic support obeying a specific parabolic scaling such that at scale $2^{-j}$, its length is $2^{-j/2}$ and its width is  $2^{-j}$. This is motivated by the parabolic scaling property of smooth curves. Other properties of the curvelet transform as well as decisive optimality results in approximation theory are reported in \citep{Curvelets-StMalo,CandesDonohoCurvelets}. Notably, curvelets provide optimally sparse representations of manifolds which are smooth away from edge singularities along smooth curves.
Several digital curvelet transforms \citep{cur:donoho99,starck:sta01_3,cur:demanet06} have been proposed which attempt to preserve the essential properties of the continuous curvelet transform and many papers \citep{starck:sta04,felix2008,starck:sta04} report on their successful application in image processing experiments. 
The so-called first generation discrete curvelet described in \citep{cur:donoho99,starck:sta01_3}  consists in applying the ridgelet transform to sub-images of a wavelet decomposition of the original image. By construction, the sub-images are well localized in space and frequency and the subsequent ridgelet transform provides the necessary directional sensitivity. This latter implementation in combination with the good geometric properties of the Healpix pixelization scheme, inspired the digital curvelet transform on the sphere~\citep{starck:sta05_2}.
%
The digital curvelet transform on the sphere is clearly invertible in the sense that each step of the overall transform is itself invertible. The curvelet transform on the sphere has a redundancy factor of $16J +1$ when $J$ scales are used, which may be a problem for handling huge data sets such as from the future Planck-Surveyor experiment. This can be reduced by substituting the pyramidal wavelet transform to the undecimated wavelet transform in the above algorithm. More details on the wavelet, ridgelet, curvelet algorithms on the sphere can be found in \citep{starck:sta05_2} and software related to these new transforms is available from the web page {\it http://jstarck.free.fr/mrs.html}. 
As for the isotropic wavelet on the sphere, a straightforward extension to polarized data will consist in applying successively the curvelet transform on the sphere to the
three components $T$, $Q$ and $U$. Figure~\ref{fig_qucur_back} shows the backprojection of a Q-curvelet coefficient and U-curvelet coefficient. Clearly,
the shapes of these polarized curvelet functions are very different from the polarized wavelet functions. In the next section, we will build very different dictionaries using the E/B mode decomposition.
\section{Polarized E/B Wavelet and E/B Curvelet}
\label{sec:pol_eb}

\subsection{Introduction}
We have seen that the generalization of the Fourier representation for polarized data on the sphere is the spin-2 spherical harmonics basis denoted $_{\pm 2}Y_{\ell m}$: 
\begin{equation} 
Q \pm i U  = \sum_{\ell, m}  { _{\pm 2}a_{\ell m}}   {_{\pm 2}Y_{\ell m} }
\end{equation} 
  
At this point, it is convenient~\citep{zalda} to introduce the two quantities denoted $E$ and $B$ which are defined on the sphere by 
\begin{eqnarray}\label{EB}
E = &  \sum_{\ell, m}   a_{\ell m} ^E Y_{\ell m} =  \sum_{\ell, m}  - \frac{ 1}{2}   ({_{ 2}a_{\ell m}}  +  {_{- 2}a_{\ell m}} )    Y_{\ell m} \\ \nonumber
B = & \sum_{\ell, m}   a_{\ell m} ^B Y_{\ell m} =  \sum_{\ell, m}  i \frac{ 1}{2}    ({_{ 2}a_{\ell m}}  -  {_{- 2}a_{\ell m}} )   Y_{\ell m} 
\end{eqnarray} 
where $Y_{\ell m}$ stands for the usual spin 0 spherical harmonics basis functions. The quantities $E$ and $B$ are derived by applying the spin lowering operator twice to $Q +  i U$  and the spin raising operator twice to $Q - i U$ so that $E$ and $B$ are real scalar fields on the sphere, invariant through rotations of the local reference frame. The normalization of $a_{\ell m} ^E$ and $a_{\ell m} ^B$ chosen in the latter definition is purely conventional but it appears to be  rather popular~\cite{1997PhRvD..55.1830Z,2003PhRvD..67b3501B}. Still, we could multiply $a_{\ell m} ^E$ and $a_{\ell m} ^B$ by some $A_{\ell}$ and we would have just as good a representation of the initial polarization maps. Through a change of parity $E$ will remain invariant whereas the sign of pseudo-scalar $B$ will change. The $E$ and $B$ modes defined here are not so different from the \emph{gradient} (\emph{i.e. curl} free) and \emph{curl} (\emph{i.e. divergence} free) components encountered in the analysis of vector fields. Finally, the spatial anisotropies of the Gaussian CMB temperature and polarization fields are completely characterized in this new linear representation by the power spectra and cross spectra of $T$, $E$ and $B$. Thanks to the different parities of $T$ and $E$ on one side and $B$ on the other, the sufficient statistics reduce to only four spectra namely $C_\ell^{EE},C_\ell^{TE}, C_\ell^{TT}, C_\ell^{BB}$. For a given cosmological model, it is possible to give a theoretical prediction of these spectra. Aiming at inverting the model and inferring the cosmological parameters, an important goal of CMB temperature and polarization data analysis is then to estimate the latter power spectra, based on sampled, noisy sometimes incomplete $T,Q,U$ spherical maps.

\subsection{E/B Isotropic Wavelet}
Following the above idea of representing CMB polarization maps by means of $E$ and $B$ modes, we propose a formal extension of the previous undecimated isotropic wavelet transform that will allow us to handle linear polarization data maps $T,Q,U$ on the sphere. Practically, the maps we consider are pixelized using for instance the Healpix pixelization scheme. In fact, we are not concerned at this point with the recovery of E and B modes from pixelized or incomplete data maps which itself is not a trivial task. The extension of the wavelet transform on the sphere we describe here makes use of the $E$ and $B$ representation of polarized maps described above in a formal way. Given polarization data maps $T$, $Q$ and $U$, the proposed wavelet transform algorithm consists of the following steps : 
\vspace{.1cm}
\begin{center}
\begin{minipage}[b]{0.85\linewidth}
\footnotesize{
\begin{enumerate}
\item Apply the spin $\pm 2$ spherical harmonics transform to $Q+iU$ and $Q-iU$. Practically, the Healpix software package provides an implementation of this transform for maps that use this pixelization scheme. Otherwise, a fast implementation was recently proposed by \citet{wiauxspin2} 
\item Combine the decomposition coefficients ${ _{2}a_{\ell m}}$ and ${ _{-2}a_{\ell m}}$ from the first step into $a_{\ell m}^E$ and $a_{\ell m}^B$ and build \emph{formal} $E$ and $B$ maps associated with $Q$ and $U$ by applying the usual inverse spherical harmonics transform, as in equation~\ref{EB}.  For numerical and algorithmic purposes, it may be efficient to stay with the spherical harmonics representation of $E$ and $B$.
\item Apply the undecimated isotropic transform on the sphere described above to map $T$ and to the $E$, $B$ representation of the polarization maps. 
\end{enumerate}}
\end{minipage}
\end{center}
\vspace{.1cm}
The wavelet coefficient maps  $w_j^T$, $w_j^E$, $w_j^B$ and the low resolution approximation maps $c_J^T$, $c_J^E$, $c_J^B$ are obtained by applying the isotropic undecimated wavelet transform described in section~\ref{sec:UWTS} to the $T$, $E$, $B$ representation of the polarized data.  Figure~\ref{fig:UWTSpol} shows the result of applying the proposed transform to the polarized CMB data map \emph{ka} \footnote{available at http://lambda.gsfc.nasa.gov/product/map/current/ } from the WMAP experiment. The top two images show the initial $Q$ and $U$ maps while the subsequent maps are the low pass and wavelet coefficients' maps in a four scale decomposition. The scaling function we used is a cubic box spline as proposed in section~\ref{sec:UWTS}. The wavelet coefficients were obtained as the difference between two successive low pass approximations of the multiresolution decomposition of the $E$ and $B$ maps. The proper choice for  the scaling and wavelet functions will depend on the application and the existence of constraints to be enforced.
\begin{figure*}
\vbox{
\centerline{
\hbox{
}
}
\centerline{
\hbox{
}
}
\centerline{
\hbox{
}
}
\centerline{
\hbox{
}
}
\centerline{
\hbox{
}
}
}
\caption{\textbf{top~:} $Q$ and $U$ CMB polarization data maps from channel \emph{ka} of the WMAP  experiment.  \textbf{left~:} low pass and wavelet coefficients in three scales of the formal E mode. \textbf{right~:} low pass and  wavelet coefficients in three scales of the formal B mode.}
\label{fig:UWTSpol}
\end{figure*}

\subsection*{Reconstruction}
\begin{figure*}[htb]
\centerline{
\vbox{
 \hbox{
 }
  \hbox{
 }
 \hbox{
 }
 }
 }
\caption{E-isotropic wavelet transform backprojection (left) and B-isotropic wavelet backprojection (right).}
\label{fig_eb_iwt_back}
\end{figure*}
Obviously, the transform described above is invertible and the inverse transform is readily obtained by applying the inverse of each of the three steps in reverse order. If, as in the example decomposition above, we take the wavelet function to be the difference between two successive low pass approximations, the third step is inverted by simply summing the last low pass approximation with the maps of wavelet coefficients from all scales as in equation~\ref{IWT} : 
\begin{eqnarray}
T  & = & c_{J}^T + \sum_{j=1}^{J} w_j^T \quad \quad \nonumber \\
E & = & c_{J}^E + \sum_{j=1}^{J} w_j^E \quad \quad \nonumber \\
B & = &  c_{J}^B + \sum_{j=1}^{J} w_j^B
\end{eqnarray}
where $c_{J}^X$ stands for the low resolution approximation to component $X$ and $w_j^X$ is the map of wavelet coefficients of that component on scale $j$. Finally, noting that : 
\begin{eqnarray}
Q  & =  & -\frac{1}{2} \sum_{\ell, m}   a_{\ell m} ^E   ( {_{ 2} Y}_{\ell m} +  {_{ -2} Y}_{\ell m} ) +  i a_{\ell m} ^B ( {_{ 2} Y}_{\ell m} -  {_{ -2} Y}_{\ell m} )  \nonumber \\
     & =  & \sum_{\ell, m}   a_{\ell m} ^E   Z_{\ell m}^+ +  i a_{\ell m} ^B Z_{\ell m}^-   \\ \nonumber
U  & =   & -\frac{1}{2} \sum_{\ell, m}   a_{\ell m} ^B   ( {_{ 2} Y}_{\ell m} +  {_{ -2} Y}_{\ell m} ) -  i a_{\ell m} ^E ( {_{ 2} Y}_{\ell m} -  {_{ -2} Y}_{\ell m} )  \nonumber \\
      & =  & \sum_{\ell, m}   a_{\ell m} ^B Z_{\ell m}^+ -  i a_{\ell m} ^E Z_{\ell m}^-   
\end{eqnarray}
the initial representation of the polarized data in terms of $T$, $Q$ and $U$ maps is reconstructed from its wavelet coefficients using the following equations : 
\begin{eqnarray}\label{eq:recons}
T =& c_{J}^T + \sum_{j=1}^{J} w_j^T \\ \nonumber
Q =& c_{J}^{E,+}  + i c_{J}^{B,-}  +   \sum_{j=1}^{J}  \Big \{  w_j^{E,+} + i w_j^{B,-}  \Big \} \\ \nonumber 
U =& c_{J}^{B,+}  -  i c_{J}^{E,-}  +   \sum_{j=1}^{J}  \Big \{  w_j^{B,+} - i w_j^{E,-}   \Big \}
\end{eqnarray}
where
\begin{eqnarray}\label{eq:change}
c_{J}^{X,+}  =  c_{J}^X  \sum_{\ell, m} Y_{\ell m}^{\dagger} Z_{\ell m}^+   \quad \textrm{and} \quad  c_{J}^{X,-}  =  c_{J}^X  \sum_{\ell, m} Y_{\ell m}^{\dagger} Z_{\ell m}^- 
\end{eqnarray}
with $W^\dagger$ denoting the transpose conjugate of $W$ so that $\tilde{W} W^\dagger$ is the scalar dot product of $\tilde{W}$ and $W$ while $W^\dagger \tilde{W}$ is an operator (or matrix) acting on its left hand side as a projection along $W$ and \emph{reconstruction} along $\tilde{W}$. In practice, the Healpix software package provides us with an implementation of the forward and inverse spin $0$ and spin $2$ spherical harmonics transforms which we need to implement the proposed inverse transform given by equations~\ref{eq:recons} and~\ref{eq:change}. Clearly, as mentioned earlier in section~\ref{sec:UWTS}, we could have chosen some other wavelet function than merely the difference between two consecutive scaling functions, and the transformation would still be nearly as simple to invert. Fig.\ref{fig_eb_iwt_back} shows, on the left, backprojections of E-wavelet coefficients, and, on the right,  backprojections of  B-wavelet coefficients on the right hand side at different scales.

\subsection*{E-B  Curvelet}
\begin{figure*}[htb]
\centerline{
\vbox{
 \hbox{
 }
  }
 }
\caption{E-curvelet coefficient backprojection.}
\label{fig_ecur_back}
\end{figure*}

\begin{figure*}[htb]
\centerline{
\vbox{
 \hbox{
 }
 }
 }
\caption{B-curvelet coefficient backprojection.}
\label{fig_bcur_back}
\end{figure*}
Similarly to the EB-wavelet constructions, we can easily construct an EB-curvelet transform by first  computing the E and B components using the spin $\pm 2$ spherical harmonics transform, and then applying a curvelet 
transform on the sphere separately on each of these two components.
Fig.~\ref{fig_ecur_back}  shows the backprojection of an E-curvelet coefficient and Fig.~\ref{fig_bcur_back} shows the backprojectionof a B-curvelet coefficient.


\begin{figure*}[htb]
\centerline{
\vbox{
 \hbox{
 }
 \hbox{
 }}}
\caption{Top, simulated input polarized image (left) and noisy polarized field (right)
Bottom, denoising of the polarized field using the EB-isotropic undecimated wavelets (left) and the EB-curvelets (right).}
\label{fig_poldust_denoising}
\end{figure*}

\section{Experiments~: Application to denoising}
\label{sec:pol_denoising}

\begin{table}
\caption{ Error (in dB) for both the $Q$ and $U$ components between the original dust image and  respectively 
its noisy version and the results obtained by hard thresholding representations of the noisy data in different dictionaries.}
\begin{tabular}{l|ccccc}\hline
Q Component   &                    &  &  & &  \\      \hline
   Noisy image   & $20.01$    &   $13.98$   &   $6.03$     &   $-0.007$   &   $-6.02$   \\      \hline \hline
QU-UWT            &  $34.91$   &   $32.41$   &   $28.88$   &   $26.41$    &   $23.68$ \\
EB-UWT            &   $33.52$   &  $30.63$   &   $27.83$   &  $25.86$     &   $23.91$   \\
QU-CUR           &   $34.14$   &  $ 31.12$     &    $28.56$   &  $26.58$  &    $24.56$   \\
EB-CUR           &    $ 33.36$   &  $30.60$   &    $28.17$   &  $26.30$    &   $ 23.99$    \\   
OWT                   &   $34.07$   & $30.77$   &    $27.08$   &   $24.57$   &    $21.33$     \\    
Mod-phase     &    $30.51$    & $26.42$  &     $20.97$   &    $16.25$  &   $11.05$  \\
\hline
U Component  &                 &  &  & &                \\      \hline
Noisy image   & $19.99$    &   $13.97$   &   $6.01$     &   $-0.001$      &   $-6.01$   \\      \hline  \hline
QU-UWT           &   $40.88$   &   $39.22$   &   $36.77$   &   $35.50$     &   $31.90$ \\
EB-UWT            &   $38.80$   &  $36.71$   &   $35.41$   &  $34.67$       &   $32.23$  \\
QU-CUR           &   $39.54$   &  $ 37.85$     &    $36.68$   &  $35.83$    & $32.38$     \\
EB-CUR            &    $39.64$   &  $ 37.72$   &    $35.33$   &  $33.51$    &  $29.30$    \\     
OWT                   &   $39.18$    & $37.29$   &    $33.30$   &   $29.06$   &    $23.56$   \\ 
Mod-phase     &    $30.76$    & $26.81$  &     $21.31$   &    $16.83$  &   $11.36$  \\  \hline
\hline
\end{tabular}\label{tab_psnr_pol_dust}
\end{table}

\begin{table}
\caption{ Error (in dB) for both the $Q$ and $U$ components between the original synchrotron image and  respectively 
its noisy version and the results obtained by the hard thresholding using different dictionaries.}
\begin{tabular}{l|ccccc}\hline
Q Component   &                    &  &  & &  \\      \hline
   Noisy image   & $19.99$    &   $13.98$   &   $6.02$     &   $-0.005$   &   $-6.02$   \\      \hline \hline
QU-UWT            &  $33.40$   &   $31.27$   &   $27.78$   &   $24.71$    &   $21.16$ \\
EB-UWT            &   $33.03$   &  $30.53$   &   $26.79$   &  $23.86$     &   $20.76$   \\
QU-CUR           &   $35.04$   &  $32.92$     &    $29.19$   &  $26.19$  &    $22.37$   \\
EB-CUR           &    $34.83$   &  $32.09$   &    $27.89$   &  $24.81$    &   $ 21.06$    \\   
OWT                   &   $33.20$    & $30.10$   &    $ 25.72$   &   $22.62$   &    $19.39$     \\    
MP-UWT     &    $26.76$    & $23.05$  &     $17.85$   &    $13.69$  &   $8.97$  \\  \hline
U Component  &                 &  &  & &                \\      \hline
Noisy image    & $19.99$    &   $13.97$   &   $6.01$     &   $-0.007$      &   $-6.01$   \\      \hline  \hline
QU-UWT           &   $32.90$   &   $31.25$   &   $29.04$   &   $27.51$     &   $25.27$ \\
EB-UWT            &   $33.22$   &  $31.43$   &   $29.11$   &  $26.99$       &   $24.35$  \\
QU-CUR           &   $33.79$   &  $ 31.75$     &    $29.28$   &  $27.71$    & $25.46$     \\
EB-CUR            &    $34.05$   &  $ 31.98$   &    $29.38$   &  $27.17$    &  $24.07$    \\     
OWT                   &   $32.69$    &    $30.49$   &   $27.87$   & $25.49$ &    $21.87$   \\ 
MP-UWT     &    $26.76$    &   $22.76$   & $17.85$  &     $13.86$   &    $ 9.41$   \\  \hline
\hline
\end{tabular}\label{tab_psnr_pol_sync}
\end{table}

Thanks to the invertibility of the different proposed transforms for polarization maps on the sphere, it is possible to use these transformations for restoration applications. 
The denoising algorithm we use here consists in three consecutive steps~:
\begin{enumerate}
\item Apply the chosen polarized multiscale transform. 
\item Set to zero those coefficients whose absolute values are below a given threshold. We have used a threshold equal to five times the noise standard deviation.
\item Reconstruct the denoised field using the inverse transform.
\end{enumerate}
More sophisticated thresholding strategies exist \citep{starck:book06} which can be used just as well on coefficients of polarized wavelets and curvelets.

Figure~\ref{fig_poldust_denoising} illustrates  the results of a simple denoising experiment. 
Noise was added to a simulated dust image (see Fig.~\ref{fig_poldust_denoising} top left and right), and the noisy QU-field was filtered  by thresholding either the EB-isotropic wavelet coefficients of the polarized dust map (Fig.~\ref{fig_poldust_denoising} bottom left)
or the EB-isotropic curvelet coefficients (Fig.~\ref{fig_poldust_denoising} bottom right). Both decompositions produce nice visual results.

In order to be more quantitative, we used two test images (synchrotron and dust) with different noise levels. The noise levels were taken equal to $0.1,0.2,0.5,1.$ and $2$ times the standard deviation of the noise-free image.
On each noisy image, we applied six different transformations, thresholded  the coefficients and reconstructed the filtered images. The transforms we used are the QU-wavelets  (QU-UWT), 
the EB-wavelets (EB-UWT), the QU-curvelet  (QU-CUR),  the EB-curvelet (EB-CUR), the biorthogonal wavelet transform (OWT) and the modulus-phase undecimated multiscale transform (MP-UWT).
For each filtered image  $Q$  (resp. $U$), we computed the Signal-to-Noise Ratio (SNR) in dB~:
\begin{equation}
E  = 10 \log_{10}(  \sigma_I^2 /  \sigma_e^2 )
\end{equation}
where $\sigma_I$ is the standard deviation  of the noise free original image component $Q$ (resp. $U$) and $\sigma_e$ is the standard deviation of the error image \textit{i.e.} the difference between the noise-free component $Q$  (resp. $U$) and the filtered component  $Q$  (resp. $U$). These errors are given in Table~(\ref{tab_psnr_pol_dust}) for the dust image and in Table~(\ref{tab_psnr_pol_sync}) for the synchrotron image.

It is clear from the above results that the different transforms described here do not perform as well on this specific numerical experiment. For the dust, QU-wavelets are better when the noise level is not so high, while curvelets do better when the noise and signal levels are of the same order. This can be explained by the fact that structures on small scales are more or less isotropic, and therefore 
better represented by wavelets, while large scale structures are anisotropic and therefore better analyzed using curvelets. When increasing the noise level, structures on the smallest scales are no longer detected by either of the two dictionaries. Only large scale features are detectable, and curvelets do this job better. Dealing with the polarized synchrotron map, curvelets  do better than wavelets in all cases experimented with here. Although the bi-orthogonal wavelet transform is clearly not as good as the others in these experiments, it could nevertheless be very useful in situations where computation time is an important issue. Indeed, since it doesn't make use of the spherical harmonics transform and also because it is not redundant, it is a very fast transform.
Finally the worse results were obtained using the Modulus-Phase multiscale transform. This can be explained by the fact that the Gaussian approximation we made for the noise in the wavelet transform of the modulus  is not accurate enough. Furthermore, it is not clear what is the best way to correct the phase wavelet coefficients from the noise. A better understanding of the noise behavior after transformation is clearly required before the Mod-phase multiscale transform can be used for restoration purposes. However, for other applications such as non-Gaussianity studies, the latter transform could prove an interesting tool to use as well.

\section{Conclusion}
 \label{sec:conc}
 
The present contribution has enlarged the set of available representations of polarized data on the sphere. We described the construction of new multiscale decompositions, namely the modulus-phase transform, the QU and EB isotropic undecimated wavelet and curvelet transforms for polarized data on the sphere. 
The latter are derived as extensions of the undecimated wavelet and curvelet transforms for scalar maps on the sphere. 
The proposed extensions use a formal representation of $T$, $Q$ and $U$  polarization data maps in terms of the scalar $T$, $E$ and pseudo-scalar $B$ maps. The proposed transforms are invertible allowing for applications in image restoration and denoising as was shown in a preliminary experiment. 
Ongoing research is concerned with assessing and understanding the merits of the different transforms we described, in such problems as restoring, denoising or inpainting of sparse linearly polarized data, as well as in the blind separation of mixed linearly polarized components.  

\noindent
 
\subsubsection*{Software}
The software related to this paper, {\bf MRS/POL}, and its full documentation 
will be included in the next version of the the  {\bf MRS} software which is
available from the following web page:
\begin{verbatim}
       http://jstarck.free.fr/mrs.html
 \end{verbatim}

\begin{acknowledgements}
Some of the results in this paper have been derived using the Healpix  \citep{pixel:healpix}.
\end{acknowledgements}

\bibliographystyle{astron} 
\bibliography{JLSBibTex}

\end{document}